\documentclass[aps,prb,preprint,showpacs]{revtex4}
\usepackage{natbib}
\usepackage{epsfig}
\usepackage{color}

\begin{document}

\title{The action of the nearest neighbor Coulomb repulsion on the homogeneity
in the high concentration domain for itinerant systems.}
\author{Zsolt~Gulacsi} 
\affiliation{Department of Theoretical Physics, University of
Debrecen, H-4010 Debrecen, Bem ter 18/B, Hungary}
\date{\today }

\begin{abstract}
  Exact results are presented for itinerant systems
  demonstrating that the nearest neighbor Coulomb
  repulsion (V) destroys the homogeneity in the high concentration regime, this
  property being not present in the low concentration domain. Since the effects
  of V often seems contradictory, and the number of phases in which it could
  appear is extremely large, this result underlines that the action of the
  nearest neighbor repulsion is not necessarily routed in the characteristics
  of phases on which it acts, but could be intimately related to V itself.
  The $V > 0$ case usually means non-integrability, hence the deduced exact
  ground states are related to non-integrable systems, the technique being
  based on positive semidefinite operator properties.
\end{abstract}

\maketitle


\section{Introduction}

The basic fermionic, itinerant and interacting models of condensed matter
physics, besides the on-site repulsion (Hubbard) term U, usually do not
contain a nonlocal nearest neighbor repulsion term V as well \cite{VV1}.
Indeed, if good quality results far beyond mean-field are intended to be
deduced, the presence of V introduces real mathematical complications (e.g.
transforms the integrable model in a non-integrable model \cite{VV2}),
hence at the beginning, V was treated as a second row spectator, used mainly 
(in the thermodynamic limit) as a supplementary contributing factor, sometimes
helping/not disturbing significantly \cite{VV3}, sometimes working against
\cite{VV4} the studied phase. On a global view, the presence of
the V term was intensively analyzed \cite{VVXX1,VVXX2,VVXX3,VVXX4,VVXX5,VVXX6},
and even exact ground states have been deduced in localized cases
\cite{VVY1,VVY2,VVY3,VVY4,VVY5}.
    
Advancing in time, with the accumulation of the data, it has turned out 
to be extremely important to analyze and understand the real physical
background of the action of V \cite{VVXX2,VVXX4}.
This is strongly underlined e.g. by the
observations stressing that the effect of V becomes even ``antagonistic''
\cite{VV5},
helping the development of the studied phase in the underdoped regime, and
working against of the studied state in the high concentration regime
\cite{VV5}. Contrary to this, sometimes it is observed that around half filling
a resilience of the studied phase to nearest neighbor repulsion appears
\cite{VV6}. The studied phases have been different, depending on the scientific
interest of the research groups describing the results, those groups trying to
explain the observed behavior based on specific characteristics of the studied
phases \cite{VV5,VV6,VV7,VV8}. However those phases have covered a large
spectrum, ranging from superconductivity \cite{VV5}, magnetism \cite{VV7} to
density waves \cite{VV8}, hence surely the observed behavior has had also a
background/a physical cause connected to V itself, independent on the phases
on which V is acting.

The presented paper concentrates for itinerant systems
on this aspect, and shows, based on exact
results, that the nearest neighbor Coulomb repulsion V, in high concentration
region destroys the homogeneity, strongly influencing in this manner the phases
present in the system. In the low concentration region, a such kind of
characteristic is not present. The presented results are
completely new, and in my knowledge have not been treated in other
publications. 
  
The used technique starts from exact ground states deduced in the high
concentration regime at $V=0$ and described in literature, and deduces in the
same conditions the exact ground state at $V > 0$, comparing both results, and
underlining that the obtained differences are not characteristic for the low
concentration domain. Since $V > 0$ usually means non-integrability (see e.g.
Ref.(\cite{VV2})), the deductions are made for non-integrable systems using
methods connected to positive semidefinite operator properties.

I would like to mention, that the process of the deduction of the
exact ground states describing itinerant non integrable systems has connections
to the treatment of non linear system of equations. This is because the
deduction of the block operators entering in the ground state expressions, is
connected to the solutions of non-linear complex algebraic system of equations
(i.e.the matching equations, see e.g.
Refs.(\cite{VVX1,VVX2,VVX3,VVX4,VVX5,VVX6,VVX7,VVX8})).
This also enrolls in the consequential
efforts made for solving and understanding non linear systems of equations
describing many-body systems present in several fields of physics
\cite{NN1,NN2,NN3,NN4,NN5}.  
  
The remaining part of the paper is structured as follows: Section II
presents in general the transcription technique of the Hamiltonian intended to
be used for the exact description of the system in the high concentration
region, Section III presents exact ground states for the diamond chain case,
Section IV describes exact ground states for two bands two dimensional lattice
with many-body spin-orbit interactions, and finally, Section V containing the
summary and discussions closes the presentation.

\section{The Hamiltonian used in the high concentration region.}

\subsection{The interaction energy terms.}

First let us prepare the interaction terms for the study in the region of high
concentration. The used interaction terms have the standard form
\begin{eqnarray}
\hat H_I = \hat H_U + \hat H_V, \quad \hat H_U = \sum_j U_j \hat n^{\uparrow}_j
\hat n^{\downarrow}_j, \quad \hat H_V = V \sum_{<k,l>} \hat n_l \hat n_j,
\label{EQ1}
\end{eqnarray}
where $j$ represents the site, $<k,l>$ denotes nearest neighbor sites taken
once into account in the sum, $U_j,V > 0$ is present, while $\hat n_j =
\hat n^{\uparrow}_j + \hat n^{\downarrow}_j$ holds, respectively. Note that the
$\hat H_I$ expression (as $\hat H=\hat H_{kin} + \hat H_I$ as well) is
concentration independent, but in the high
concentration region, transcribing it in another form, becomes mathematically
much more suitable for study. I also underline that the presented study uses
periodic boundary conditions, and in order to maintain a reasonable
extent, only those characteristics are highlighted, which are necessary for
the understanding of the paper's message.

First, to be more precise, we modify the site notation. In general a periodic
system is build up from cells, and in-cell atoms, hence one considers
$j=(i,n)$. Here $i$ denotes the cells (whose number is considered $N_c$).
Furthermore, $n$ is an in-cell numbering of sites. Hence, the in-cell atoms
are placed for a fixed cell at the positions $\vec r_n$ (where $n =1,2,..N_a$
holds). Consequently, the number of sites in the system becomes $N_s=N_c N_a$.

Second, since in the case of an exact treatment in the high concentration
limit, $\hat H_U$ and $\hat H_V$ can be identically transformed in a
mathematically much more suitable form, we perform these transformations.
One start with $\hat H_U$ which becomes:
\begin{eqnarray}
\hat H_U &=& \sum_i \sum_n U_n \hat n^{\uparrow}_{i,n} \hat n^{\downarrow}_{i,n}
=\sum_{i,n} U_n [\hat n^{\uparrow}_{i,n} \hat n^{\downarrow}_{i,n} -(
\hat n^{\uparrow}_{i,n} + \hat n^{\downarrow}_{i,n}) +1] + \sum_{i,n} U_n (
\hat n^{\uparrow}_{i,n} + \hat n^{\downarrow}_{i,n})- \sum_{i,n}U_n
\nonumber\\
&=& \sum_{i,n} U_n \hat P_{i,n} + \sum_{i,n,\sigma} U_n \hat n^{\sigma}_{i,n}
- N_c\sum_n U_n.
\label{EQ2}
\end{eqnarray}
Here one considers $U_{i,n}=U_n$ because of the periodicity of the system, and
\begin{eqnarray}
\hat P_{i,n} = \hat n^{\uparrow}_{i,n} \hat n^{\downarrow}_{i,n} -(
\hat n^{\uparrow}_{i,n} + \hat n^{\downarrow}_{i,n}) +1.
\label{EQ3}
\end{eqnarray}
I further note that if one considers $U_n=U$, so all sites have the same
on-site Coulomb repulsion, then (\ref{EQ2}) becomes $\hat H_U =U \sum_j
\hat P_j + U (N_e - N_s)$, where $N_e$ denotes the total number of electrons.
Since the second term in the last line of (\ref{EQ2}) renormalizes the on-site
one particle potentials, and the third term is a scalar, all effects of
$\hat H_U$ appear in the $\hat P_{i,n}$ operators. These are positive
semidefinite operators, which attain their minimum possible eigenvalue zero,
when all sites of the system are at least singly occupied. During an exact
treatment procedure at high concentration of carriers, it is much more easy
to analyze the effects of $\hat H_U$ via (\ref{EQ3}), instead of the
$U_n \hat n^{\uparrow}_{i,n} \hat n^{\downarrow}_{i,n}$ form present in the original
expression from (\ref{EQ1}). This information is not a novelty, and it was
several times applied during exact results deduction for non-integrable systems
\cite{VVX1,VVX2,VVX3} in the past.

Now one concentrates on $\hat H_V$, whose transformation, contrary to the case
of $\hat H_U$ described previously for completeness, is presented here for the
first time. One has
\begin{eqnarray}
\hat H_V &=&V \sum_{<k,l>} \hat n_k \hat n_l =V \sum_{<k,l>}[\hat n_k \hat n_l -2
(\hat n_k + \hat n_l) + 4] + 2V \sum_{<k,l>} (\hat n_k + \hat n_l) -4
\sum_{<k,l>} V
\nonumber\\
&=& V \sum_{<k,l>} \hat R_{<k,l>} + 2V \sum_{<k,l>} (\hat n_k + \hat n_l) -
4b_c N_c V,
\label{EQ4}
\end{eqnarray}
where
\begin{eqnarray} 
\hat R_{<k,l>} = \hat n_k \hat n_l -2(\hat n_k + \hat n_l) + 4,
\label{EQ5}  
\end{eqnarray}
and in the last term, since $\sum_{<k,l>}$ is a sum over bonds, so if one
denotes by $b_c$ the number of bonds per cell, then $\sum_{<k,l>} = b_c \sum_i=
b_c N_c$ holds. From the result obtained in (\ref{EQ4}), the second term on the
second line must be further analyzed. It can be checked that denoting by $q_n$
how many times a given site n from a cell appears in $\sum_{<k,l>}$, one finds
that $\sum_{<k,l>} (\hat n_k + \hat n_l) = \sum_i \sum_n q_n \hat n_{i,n}$ holds.
Consequently the transformation expression of $\hat H_V$ becomes
\begin{eqnarray}
\hat H_V = V \sum_{<k,l>} \hat R_{<k,l>} +2V \sum_i \sum_n q_n \hat n_{i,n}
-4b_cN_c V.
\label{EQ6}
\end{eqnarray}
Similarly to the result obtained in (\ref{EQ2}), the second term in (\ref{EQ6})
renormalizes the one particle on-site potentials, the last term is a scalar,
hence all effects of $\hat H_V$ are collected by the $\hat R_{<k,l>}$ operators
which are mathematically more suitable for use at high carrier concentration
in a non-approximated treatment. The $\hat R_{<k,l>}$ operators
are again positive semidefinite operators. Their minimum eigenvalue zero
excepting one case, is attained always when $\hat P_{i,n}$ presents its minimum
eigenvalue zero, the exception being the situation when the nearest neighbor
sites $<k,l>$ are both singly occupied. Because of this reason, all
$\hat P_{i,n}$, and all $\hat R_{<k,l>}$ positive semidefinite operators attain
simultaneously their minimum eigenvalue zero when a) all sites are at least
singly occupied, and b) there are not present nearest neighbor site pairs both
singly occupied. It is important to note, that this last requirement b) makes
$\hat H_V$ the enemy of the homogeneity at high concentration of carriers. This
should be such understood that in the presence of $\hat H_V$, the nearest
neighbor sites $k,l$ will be different, hence also in average will be different.

At this stage the reader could say that also $\hat H_V$ in its initial form from
(\ref{EQ1}) punishes the system at singly occupied nearest neighbor sites
(i.e. $n_k=1,n_l=1$), then why the effect appears in the high concentration
region only ? The answer to this question becomes to be evident when one
realizes
that $\hat H_V$ from (\ref{EQ1}) also punishes the system at $n_k=1, n_l=2$ or
$n_k=2, n_l=1$ so in the low concentration case, its effect is not concentrated
on the deletion of the homogeneity. Contrary to this, $\hat R_{<k,l>}$ from
(\ref{EQ5}) punishes the system only in the $n_k=1,n_l=1$ case ($n_k=1, n_l=2$
or $n_k=2, n_l=1$ provide the minimum eigenvalue zero). This is why the effect
becomes to be accentuated only in the high concentration region.

At this step I must mention, that deducing the ground states for
itinerant non integrable systems, given by the kinetic energy term, different
decompositions of the Hamiltonian in positive semidefinite form must be used
\cite{VVX1,VVX2,VVX3,VVX4,VVX5,VVX6,VVX7,VVX8} for high and low concentration
regions. In the high concentration region, from the V term, the $\hat R_{<k,l>}$
positive semidefinite operator enters in the transformed Hamiltonian. But
$\hat R_{<k,l>}$ is not suitable for the low concentration domain, because for
$n_k,n_l \leq 1$ which often emerge at low concentration, provides positive
eigenvalues, hence expel the system outside of the ground state.
In the small concentration region, from the V term, the positive semidefinite
operator $\hat P_1(k,l)=\hat n_k \hat n_l$ enters in the transformed positive
semidefinite Hamiltonian which provides the ground state, where
$\hat H_V = V \sum_{<k,l>} \hat P_1(k,l)$ holds. But as mentioned above,
contrary to $\hat R_{<k,l>}$, $\hat P_1(k,l)$ hence $\hat H_V$, punishes the
$n_k=1,n_l=2$ local configuration.   

\subsection{The kinetic energy term.}

A kinetic energy term $\hat H_{kin}$ is a one-particle type of Hamiltonian
which is a linear combination of $\hat c^{\dagger}_{i,n,\sigma} \hat c_{i',n',\sigma'}$
type of contributions, i.e.
\begin{eqnarray}
\hat H_{kin}= \sum_{i,i',n,n',\sigma'}
\Lambda^{i',n',\sigma}_{i,n,\sigma} \: \hat c^{\dagger}_{i,n,\sigma} \hat c_{i',n',\sigma'},
\label{EQ7}
\end{eqnarray}
where for generality, also spin-flip type of hoppings were included, and
$\Lambda^{i',n',\sigma}_{i,n,\sigma}$ represents different type of one-particle
matrix elements as hopping, hybridization, on-site one particle potentials,
etc. The kinetic Hamiltonian from (\ref{EQ7}) can be transformed as follows:
We define block operators in each cell $\hat A_{i,z,\sigma}$, where $z=1,2,..N_z$,
and often, but not obligatory, $N_z \leq N_a$, as
\begin{eqnarray}
\hat A_{i,z,\sigma} = a_{z,1} \hat c_{i,n_1,\sigma} + a_{z,2}\hat c_{i,n_2,\sigma} +
a_{z,3}\hat c_{i,n_3,\sigma} + ... + a_{z,r_z} \hat c_{i,n_{r_z},\sigma},  
\label{EQ8}
\end{eqnarray}
where $r_z$ denotes the number of components of the block operator
$\hat A_{i,z,\sigma}$, $a_{z,l}$ are numerical prefactors, and $n_l$ are on-cell
positions (hence $r_z \leq N_a$). [In order to exemplify, for example in the
case of the pentagon chain analyzed in Ref.(\cite{VVX4}) one has $N_a=6$,
$N_z =5$ (see Eq.(2)), $r_1=4, r_2=3, r_3=4, r_4=2, r_5=4$ hold, respectively].
Using (\ref{EQ8}), $\hat H_{kin}$ can be written as
\begin{eqnarray}
- \hat H_{kin} = \sum_{i,z,\sigma} \hat A^{\dagger}_{i,z,\sigma} \hat A_{i,z,\sigma}
+ \lambda,
\label{EQ9}
\end{eqnarray}
where $\lambda$ is a scalar. The block operators (\ref{EQ8}), are fermionic
operators, but not canonical Fermi operators, and have the
following main properties
\begin{eqnarray}
\hat A^{\dagger}_{i,z,\sigma} \hat A^{\dagger}_{i,z,\sigma} = 0, \quad
\{\hat A_{i,z,\sigma}, \hat A^{\dagger}_{i,z,\sigma} \} = \sum_{j=1}^{r_z} |a_{z,j}|^2,
\quad \hat A^{\dagger}_{i,z,\sigma} \hat A^{\dagger}_{i',z',\sigma'} = -
\hat A^{\dagger}_{i',z',\sigma'} \hat A^{\dagger}_{i,z,\sigma} 
\label{EQ10}
\end{eqnarray}
Using the second property of (\ref{EQ10}), the kinetic Hamiltonian from
(\ref{EQ9}) becomes
\begin{eqnarray}
\hat H_{kin} = \sum_{i,z,\sigma} \hat A_{i,z,\sigma} \hat A^{\dagger}_{i,z,\sigma}
-\sum_{j=1}^{r_z} |a_{z,j}|^2 - \lambda = \sum_{i,z,\sigma} \hat A_{i,z,\sigma}
\hat A^{\dagger}_{i,z,\sigma} + C
\label{EQ11}
\end{eqnarray}
where it is important to underline that the first term is a positive
semidefinite operator, while $C=-\sum_{j=1}^{r_z} |a_{z,j}|^2 - \lambda$ is a
scalar. This is the kinetic Hamiltonian form which
mathematically is most suitable for use in the high concentration region.

The reason for this is connected to the fact that the construction of the
ground state corresponding to the Hamiltonian $\hat H = \hat H_{kin} + \hat H_I$
in the case of (\ref{EQ11}) starts by the construction of the wave vector
\begin{eqnarray}
|\Psi\rangle = \prod_{i,z,\sigma}\hat A^{\dagger}_{i,z,\sigma} |0\rangle
\label{EQ12}  
\end{eqnarray}
where $|0\rangle$ is the bare vacuum with no fermions present.
Since given by the first property of (\ref{EQ10}), this $|\Psi\rangle$ is in
the kernel of the positive semidefinite operator from $\hat H_{kin}$
(i.e. $\sum_{i,z,\sigma} \hat A_{i,z,\sigma} \hat A^{\dagger}_{i,z,\sigma} |\Psi\rangle
=0$ holds). Hence the
construction of the ground state of $\hat H$ reduces to the introduction of
$|\Psi\rangle$ from (\ref{EQ12}) in the kernel of $\hat H_I$. Now taking into
account that $\hat A^{\dagger}_{i,z,\sigma}$ introduces one electron into the
system, counting the number of electrons from $|\Psi\rangle$ one finds $2zN_c$.
The introduction of (\ref{EQ12}) in the kernel of $\hat H_I$ adds to 
$|\Psi\rangle$ again at least $N_c$ electrons (at least one per cell). Hence
the total number of electrons in the ground state will be of order
$N_g=(2 N_z +1)N_c$, which exceeds usually considerably the
system half filling concentration $(N=N_aN_c=N_s)$.

I must also note that all Hamiltonian parameters enter in the
block operator parameters $a_{z,m}$, (see (\ref{EQ8})) via the matching equations
which, in the deduction of the block operator parameters fixes that the
starting Hamiltonian and the transformed Hamiltonian are identical. Since the
$a_{z,m}$ prefactors enter also in the $C$ value (see (\ref{EQ11})), also the
parameter $C$ contains the Hamiltonian parameters. Since the concrete
expressions of the $a_{z,m}$ coefficients are different in different models,
and here itinerant model independent conclusions are deduced, the $a_{z,m}$
values are not detailed. But the interested reader, in different concrete cases
can find the block operator parameter expressions
\cite{VVX1,VVX2,VVX3,VVX4,VVX5,VVX6,VVX7,VVX8}.    

\section{The case of the diamond chain}

\subsection{The Hamiltonian}

The diamond chain (known also as square chain, or tetragonal chain), has been
intensively studied in the past \cite{VVX2,VVX5}, so one has sufficient data to
exemplify the described effect at exact level (for the system see e.g. Fig.1.of
Ref.(\cite{VVX5}) reproduced here also as Fig.1). 
\begin{figure}[h]
\includegraphics[height=3.5cm, width=8.5cm]{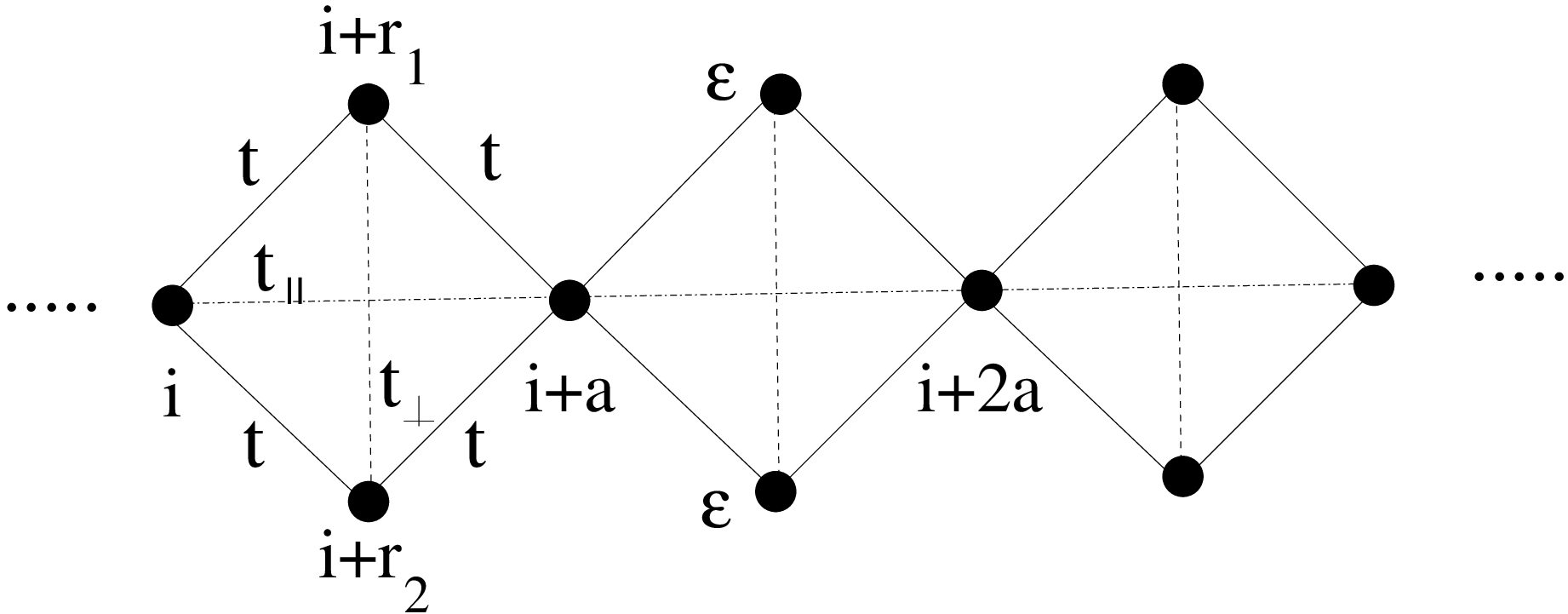}
\caption{The diamond chain. ${\bf i}$ denotes the lattice site (and the cell
origin), ${\bf a}$ is the lattice constant, $t$ and $t_{\alpha}$ with
$\alpha$ denoting the parallel and perpendicular components are hopping
matrix elements, $\epsilon$ represents on-site one particle potential,
${\bf r}_1,{\bf r}_2$, and ${\bf r}_3=0$
are the three on-cell positions.}
\end{figure}
The kinetic energy part of the Hamiltonian contains hopping along the bonds
present in the system (see Fig.1) and on-site one particle energies. Only one
block operator is needed for transforming $\hat H_{kin}$ in the form of
(\ref{EQ11}), hence $N_z=1$, the block operator being
\begin{eqnarray}
\hat A_{i,\sigma}= a_1 \hat c_{i,\sigma} + a_2 \hat c_{i+r_2,\sigma} +
a_3 \hat c_{i+a,\sigma} + a_4 \hat c_{i+r_1,\sigma}, 
\label{EQ13}
\end{eqnarray}
whose number of components $r_{z=1}=4$. One has three sites per cell, so $N_a=3$,
and the total Hamiltonian $\hat H = \hat H_{kin} + \hat H_U + \hat H_V$, after
transformation (see (\ref{EQ2}, \ref{EQ6}, \ref{EQ11})) becomes of the form
\begin{eqnarray}
\hat H = \sum_{i,\sigma} \hat A_{i,\sigma} \hat A^{\dagger}_{i,\sigma} + U \hat P +
V \hat R + C
\label{EQ14}
\end{eqnarray}
where $\hat P = \sum_{i,n}U_n \hat P_{i,n} = U \sum_j \hat P_j$,  $j=(i,n)$,
and $\hat R = \sum_{<k,l>} \hat R_{<k,l>}$, while $C$ is a scalar.

\subsection{The V=0 ground state}

The Hamiltonian
from (\ref{EQ14}) at V=0 has been analyzed in details in Ref.(\cite{VVX5}).
The maximum number of electrons in the system is $N_M = 6N_c$, and 
Ref.(\cite{VVX5}) presents an exact ground state at total number of electrons
$N=5N_c$, so concentration $\rho=N/N_M = 5/6$, which represents a correlated
half metallic state at $V=0$, possessing the ground state wave function
\begin{eqnarray}
|\Psi_g(5N_c)\rangle =(\prod_{\alpha=1}^{N_c} \hat c^{\dagger}_{n_{\alpha},
{\bf k}_{\alpha},-\sigma})(\prod_{i=1}^{N_c} [\hat c^{\dagger}_{i+r_{s_i,1},\sigma}
\hat c^{\dagger}_{i+r_{s_i,2},\sigma}]) (\prod_{i=1}^{N_c} \hat A^{\dagger}_{i,\sigma}
\hat A^{\dagger}_{i,-\sigma}) |0\rangle   
\label{EQ15}
\end{eqnarray}
see Eq.(3.17) of Ref.(\cite{VVX5}). Indeed, $\hat H -C$ from (\ref{EQ14}) is a
positive semidefinite operator, so its minimum possible eigenvalue is zero.
Than, for the first term of $\hat H$ from (\ref{EQ14}), as explained at
(\ref{EQ12}), one obtains
\begin{eqnarray}
|\Psi_1\rangle =\prod_{i=1}^{N_c} (\hat A^{\dagger}_{i,\sigma}
\hat A^{\dagger}_{i,-\sigma}) |0\rangle, \quad
(\sum_{i,\sigma} \hat A_{i,\sigma} \hat A^{\dagger}_{i,\sigma})|\Psi_1\rangle =0,
\label{EQ16}
\end{eqnarray}
and one has $2N_c$ electrons in $|\Psi_1\rangle$, $N_c$ with spin $\sigma$, and
$N_c$ with spin $-\sigma$. Now we introduce $|\Psi_1\rangle$ also in the kernel
of $\hat P$ as follows
\begin{eqnarray}
\hat F^{\dagger}_{\sigma} =\prod_{i=1}^{N_c} [\hat c^{\dagger}_{i+r_{s_i,1},\sigma}
\hat c^{\dagger}_{i+r_{s_i,2},\sigma}], \quad |\Psi_2\rangle = \hat F^{\dagger}_{\sigma}
|\Psi_1\rangle.
\label{EQ17}
\end{eqnarray}
Here $\hat F^{\dagger}_{\sigma}$ introduces in each cell, in an arbitrary in-cell
position, two electrons with spin $\sigma$. Consequently $|\Psi_2\rangle$ will
have besides $N_c$ electrons with spin $-\sigma$, also $3N_c$ electrons with
spin $\sigma$, the number of sites in the system being also $3N_c=N_aN_c=N_s$.
Hence on each site in $|\Psi_2\rangle$ one has surely one electron, the
requirements for the minimum eigenvalue of $\hat P$ are satisfied, and one
obtains
\begin{eqnarray}
\hat P |\Psi_2\rangle =0, \quad
(\sum_{i,\sigma} \hat A_{i,\sigma} \hat A^{\dagger}_{i,\sigma})|\Psi_2\rangle =0,
\label{EQ18}
\end{eqnarray}
where the last relation holds because $\hat F^{\dagger}_{\sigma} (\prod_{i=1}^{N_c}
(\hat A^{\dagger}_{i,\sigma}\hat A^{\dagger}_{i,-\sigma}) = \prod_{i=1}^{N_c}
(\hat A^{\dagger}_{i,\sigma}\hat A^{\dagger}_{i,-\sigma}) \hat F^{\dagger}_{\sigma}$ holds.
Also note that $|\Psi_2\rangle$ has $4N_c$ electrons in the system, and becomes
a ground state for $4N_c$ electrons, i.e. $|\Psi_g(4N_c)\rangle =
|\Psi_2\rangle$. But as demonstrated in Ref.(\cite{VVX5}), besides the fact that
(in $|\Psi_2\rangle$) the $\sigma$ spin electrons are localized, the $-\sigma$
spin electrons are spatially extended, but localized in the $N_c \to \infty$
thermodynamic limit.

Now one introduces further $N_c$ electrons in the system via
\begin{eqnarray} 
\hat G^{\dagger}_{-\sigma} =\prod_{\alpha=1}^{N_c} (\hat c^{\dagger}_{n_{\alpha},
{\bf k}_{\alpha},-\sigma}), \quad
|\Psi_3\rangle = \hat G^{\dagger}_{-\sigma} |\Psi_2\rangle,
\label{EQ19}
\end{eqnarray}
where $\hat G^{\dagger}_{-\sigma}$ introduces $N_c$ electrons with spin $-\sigma$
in arbitrary (available) ${\bf k}_{\alpha}$ states of the system. Since $\hat
G^{\dagger}_{-\sigma}$ operator contains only a product of creation operators, and
all creation operators anticommute, for $|\Psi_3\rangle$ the results from
(\ref{EQ18}) remain valid
\begin{eqnarray}
(\sum_{i,\sigma} \hat A_{i,\sigma} \hat A^{\dagger}_{i,\sigma})|\Psi_3\rangle =0,
\quad \hat P |\Psi_3\rangle =0,
\label{EQ20}
\end{eqnarray}
hence $|\Psi_g(5N_c)\rangle = |\Psi_3\rangle$ will be the ground state at
$N=5N_c$. As shown in Ref.(\cite{VVX5}) in this ground state the $-\sigma$
spin electrons are mobile.

I have repeated the presentation of this deduced ground state in order to have
a clear comparison possibility to the $V > 0$ solution. As can be seen, the
ground state (\ref{EQ15}) preserves the same type of properties for each site of
the system, so from this point of view can be considered ``homogeneous''

\subsection{The $V > 0$ ground state}

Now we turn back to the Hamiltonian (\ref{EQ14}), but now considering $V > 0$
as well.

The $|\Psi_1\rangle$ construction presented in (\ref{EQ16}) remains unaltered,
since involves only the kinetic term of the Hamiltonian. But $|\Psi_2\rangle$,
besides satisfying for $\hat P$ the relation (\ref{EQ18}), is not the proper
choice for $\hat R$, since contains nearest neighbor sites singly occupied.
In order to introduce $|\Psi_2\rangle$ also in the kernel of $\hat R$, the
$\hat F^{\dagger}_{\sigma}$ operator must be multiplied by an
$\hat {F'}^{\dagger}_{-\sigma}$ operator that introduces in each cell a $-\sigma$
electron at the site $i$
\begin{eqnarray}
  \hat F_1 =\hat {F'}^{\dagger}_{-\sigma}\hat F^{\dagger}_{\sigma} =
  (\prod_{i=1}^{N_c} \hat c^{\dagger}_{i,-\sigma}) \hat F^{\dagger}_{\sigma},
\quad |\Psi^V_2\rangle = \hat F_1 |\Psi_1\rangle
\label{EQ21}
\end{eqnarray}
The wave vector  $|\Psi^V_2\rangle$ contains $5N_c$ electrons. From these,
$3N_c$ electrons have spin $\sigma$, and since the number of sites is also
$N_s=3N_c$, these cannot move, hence are completely localized. Besides these,
there are also $2N_c$ electrons in the system with spin $-\sigma$. A number of
$N_c$ electrons from these are placed in the contact points between nearest
neighbor cells, hence these sites are all doubly occupied. The remaining $N_c$
number of $-\sigma$ electrons are placed by the $\hat A^{\dagger}_{i,-\sigma}$
operators from $|\Psi_1\rangle$, one in each cell, randomly in the remained
available $i+r_1,i+r_2$ in-cell positions. It can be checked that in these
circumstances
\begin{eqnarray}
(\sum_{i,\sigma} \hat A_{i,\sigma} \hat A^{\dagger}_{i,\sigma})|\Psi^V_2\rangle =0,
\quad \hat P |\Psi^V_2\rangle =0, \quad \hat R |\Psi^V_2\rangle =0,
\label{EQ22}
\end{eqnarray}
hence the ground state in presence of $V > 0$ becomes
\begin{eqnarray}
|\Psi^V_g(5N_c)\rangle = |\Psi^V_2\rangle =
(\prod_{i=1}^{N_c} \hat c^{\dagger}_{i,-\sigma})
(\prod_{i=1}^{N_c} [\hat c^{\dagger}_{i+r_{s_i,1},\sigma}
\hat c^{\dagger}_{i+r_{s_i,2},\sigma}]) (\prod_{i=1}^{N_c} \hat A^{\dagger}_{i,\sigma}
\hat A^{\dagger}_{i,-\sigma}) |0\rangle  
\label{EQ23}
\end{eqnarray}
I note that other choices for $\hat F_1$ not introduce $|\Psi_1\rangle$ in both
kernels of $\hat P$, and $\hat R$, hence $|\Psi^V_g(5N_c)\rangle$ is unique in
the conditions of the problem.

As seen, contrary to the $V=0$ ground state (\ref{EQ15}), the $V >0$ ground
state (\ref{EQ23}) is not homogeneous from the point of view of sites. The
lattice sites $i$ (in cell notation $r_3=0$) are always doubly occupied, while
the other in-cell sites not. Hence the system at $V > 0$ has periodic charge
modulation [CDW (charge density wave) type behavior], which is absent in
(\ref{EQ15}).

\section{The case of the two bands 2D square lattice with spin-orbit
interactions}

\subsection{The Hamiltonian}

The following example that I am going to present is related to a 2D strongly
correlated two-bands system with many-body spin-orbit interaction (SOI) analyzed
first in Ref.(\cite{VVX6}), used afterwards in explaining the ferromagnetism
of gold nanoclusters in Ref.(\cite{VVX7}). Also in this case there are
sufficient results present in order to allow the exemplification of the
$V \ne 0$ effects in an exact (now two dimensional) ground state.

The model contains two bands (let call them $s$ and $d$), the s-band being
non-correlated, while the d-band is correlated, containing the $\hat H_U$ and
$\hat H_V$ contributions. The unit cell of the system is seen in Fig.2.
\begin{figure}[h]
\includegraphics[height=4cm, width=4cm]{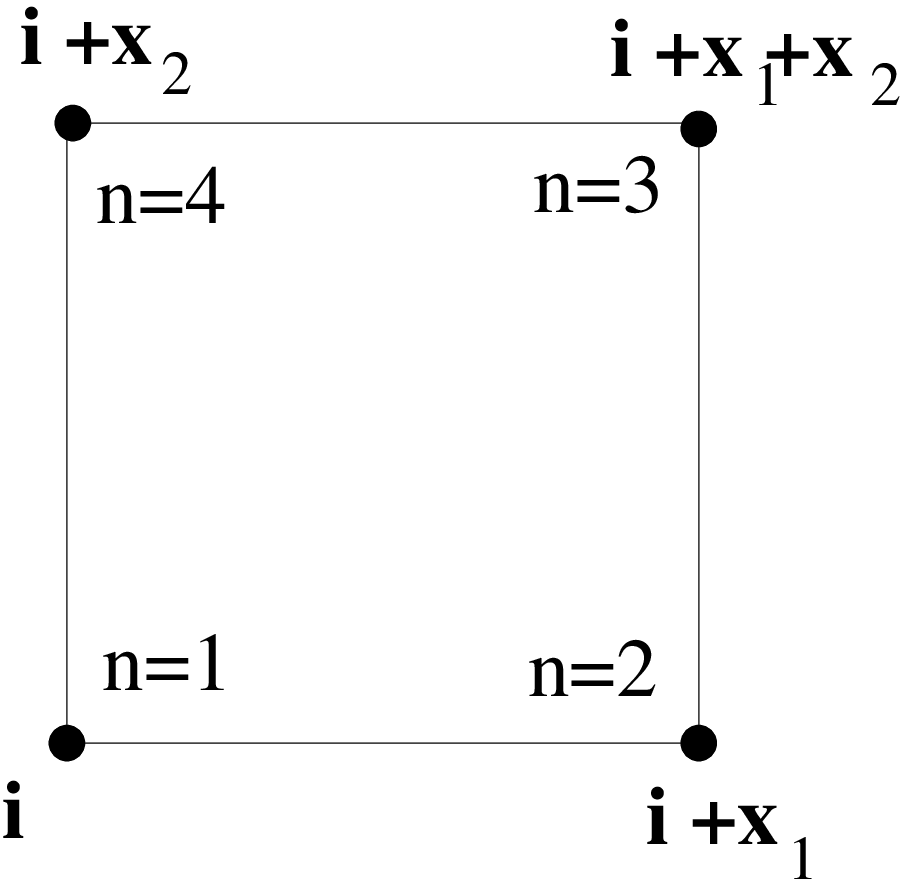}
\caption{Unit cell at lattice site ${\bf i}$, with in cell notation of sites
$n=1,2,3,4$, the Bravais vectors being denoted by ${\bf x}_1$, ${\bf x}_2$.}
\end{figure}
$\hat H_{kin}$ presented in details in Ref.(\cite{VVX6}) contains nearest
neighbor hopping terms for both bands, local and nearest neighbor hybridization
terms, on-site one particle potentials, and SOI contributions leading to
spin-flip type of hoppings. One has $N_a=1$, and two block operators are used
for the transformation (\ref{EQ9}), namely $N_z=2$, defined now with both spin
projection in order to coopt also the spin-flip contributions
\begin{eqnarray}
\hat A_{i,z} = \sum_n \sum_{\sigma}\sum_{p=s,d} a_{z,n,p,\sigma} \hat c_{p,i+r_n,
\sigma},
\label{EQ24}
\end{eqnarray}
where $\hat c_{p=1,j,\sigma}=\hat c_{j,\sigma}$ describes the non-correlated band,
and $\hat c_{p=2,j,\sigma}= \hat d_{j,\sigma}$ describes the correlated band,
and one has $r_{z=1}=r_{z=2}=16$. The interaction part of the Hamiltonian is
composed from the terms
\begin{eqnarray}
\hat H_U = \hat H^d_U = U \sum_i \hat n^d_{i,\uparrow} \hat n^d_{i,\downarrow}, \quad
\hat H_V = \hat H^d_V = V \sum_{<k,l>} \hat n^d_{k} \hat n^d_{l}.
\label{EQ25}
\end{eqnarray}
The transformed Hamiltonian becomes
\begin{eqnarray}
\hat H = \sum_{i=1}^{N_c} \sum_{z=1,2} \hat A_{i,z} \hat A^{\dagger}_{i,z} +
U \hat P + V \hat R + C,
\label{EQ26}
\end{eqnarray}
where as previously, C is a scalar [see (5-7) of Ref.(\cite{VVX6}), or (4) of
Ref.(\cite{VVX7})].  

\subsection{The ground state at $V=0$}

The system has $N_c$ sites, the maximum possible number of electrons in the
system is $N_M=4N_c$, and we concentrate now on the ground
state at number of electrons $N= 7/2 N_c$. The $V=0$ ground state has the form
\begin{eqnarray}
|\Psi_g(3.5N_c)\rangle =(\prod_{\alpha=1}^{N_c/2} \hat c^{\dagger}_{p_{\alpha},
{\bf k}_{\alpha},\sigma_{\alpha}})(\prod_{i=1}^{N_c} [\hat d^{\dagger}_{i,\uparrow} +
\hat d^{\dagger}_{i,\downarrow}]) (\prod_{i=1}^{N_c} \hat A^{\dagger}_{i,z=1}
\hat A^{\dagger}_{i,z=2}) |0\rangle,   
\label{EQ27}
\end{eqnarray}
see (8) of Ref.(\cite{VVX7}). Let us analyze (\ref{EQ27}). In constructing
$|\Psi_1\rangle$, similar to (\ref{EQ16}) one has
\begin{eqnarray}
|\Psi_1\rangle =\prod_{i=1}^{N_c}\prod_{z=1,2}\hat A^{\dagger}_{i,z}|0\rangle, \quad
( \sum_{i=1}^{N_c} \sum_{z=1,2} \hat A_{i,z} \hat A^{\dagger}_{i,z})|\Psi_1\rangle =0. 
\label{EQ28}
\end{eqnarray}
Consequently $|\Psi_1\rangle$ is inside the kernel of the first operator from
(\ref{EQ26}), and note that it contains $2N_c$ electrons. Now, introducing the
operator $\hat D^{\dagger}_i=(\hat d^{\dagger}_{i,\uparrow} +
\hat d^{\dagger}_{i,\downarrow})$ one obtains
\begin{eqnarray}
\hat D^{\dagger} = \prod_{i=1}^{N_c} \hat D^{\dagger}_i, \quad
|\Psi_3\rangle = \hat D^{\dagger} |\Psi_1\rangle.
\label{EQ29}
\end{eqnarray}
Since $|\Psi_3\rangle$ introduces on each site one $d$ electrons (with
arbitrary spin, this is the most general case), the
requirement for the minimum eigenvalue of the $\hat P$ operator is satisfied,
and $\hat D^{\dagger} (\prod_{i=1}^{N_c} \prod_{z=1,2} \hat A^{\dagger}_{i,z}) =
(\prod_{i=1}^{N_c} \prod_{z=1,2} \hat A^{\dagger}_{i,z})\hat D^{\dagger}$, hence
one has
\begin{eqnarray}  
(\sum_{i=1}^{N_c} \sum_{z=1,2} \hat A_{i,z} \hat A^{\dagger}_{i,z})|\Psi_2\rangle =0,
\quad \hat P|\Psi_2\rangle =0. 
\label{EQ30}
\end{eqnarray}
The wave function $|\Psi_2\rangle$ has $3N_c$ electrons (it is the ground state
at $N=3N_c$ which is ferromagnetic given by SOI, see Ref.(\cite{VVX6})). To
these $3N_c$ electrons one adds further $N_c/2$ new electrons via
\begin{eqnarray}
\hat G^{\dagger}_{s,d} = \prod_{\alpha=1}^{N_c/2} \hat c^{\dagger}_{p_{\alpha},
{\bf k}_{\alpha},\sigma_{\alpha}}, \quad
|\Psi_3\rangle = \hat G^{\dagger}_{s,d}|\Psi_2\rangle
\label{EQ31}
\end{eqnarray}
where $\hat G^{\dagger}_{s,d}$ introduces in the system $N_c/2$ electrons (s or d,
does not matter), with arbitrary spin, in arbitrary (available)
${\bf k}_{\alpha}$ states. Since $\hat G^{\dagger}_{s,d}$ contains only creation
operators, the properties in (\ref{EQ30}) are not altered, hence
\begin{eqnarray}  
(\sum_{i=1}^{N_c} \sum_{z=1,2} \hat A_{i,z} \hat A^{\dagger}_{i,z})|\Psi_3\rangle =0,
\quad \hat P|\Psi_3\rangle =0, 
\label{EQ32}
\end{eqnarray}
holds. The ground state at $3.5N_c$ and $V=0$ then becomes
$|\Psi_g(3.5N_c)\rangle =|\Psi_3\rangle$, so Eq.(\ref{EQ27}) is demonstrated.
As seen, the state in (\ref{EQ27}) is homogeneous, since all sites of the system
behave the same when $|\Psi_g(3.5N_c)\rangle$ is present.

\subsection{The ground state at $V > 0.$}

Since it is clear that both $|\Psi_2\rangle$ and $|\Psi_3\rangle$ may have
singly occupied nearest neighbor site pairs, the problem which should be
resolved at this stage is to introduce $|\Psi_2\rangle$ also in the kernel
of the $\hat R$ operator from (\ref{EQ26}). For this, we divide the lattice in
two sublattices (say A, and B), see Fig.3, and we
define
\begin{figure}[h]
\includegraphics[height=5cm, width=8cm]{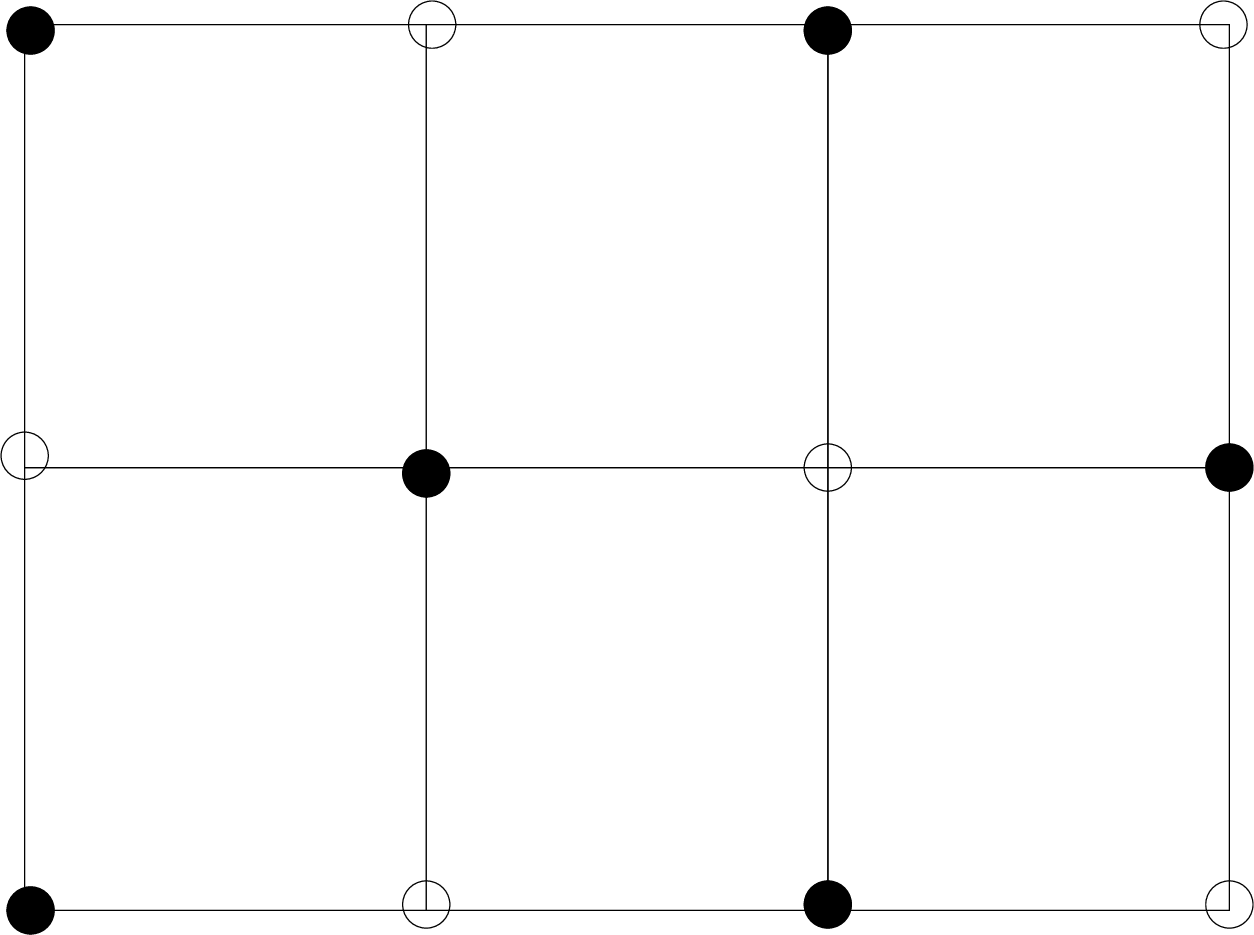}
\caption{Division of the system in two sublattices: A (open dots), and B
(black dots).}
\end{figure}
\begin{eqnarray}
  \hat D^{\dagger}_1 = \prod_{i \in A} \hat D^{\dagger}_i, \quad
  |\Psi^V_3\rangle = \hat D^{\dagger}_1 |\Psi_2\rangle,
\label{EQ33}  
\end{eqnarray}
where $\hat D^{\dagger}_1$ introduces one more electron $d$ with arbitrary spin
in one sublattice (say A). Then, the sublattice A will have double $d$ electron
occupancy on each site, while the sublattice B (at least) single $d$ electron
occupancy on
each site. In this manner the requirements for the minimal eigenvalue of both
$\hat P$ and $\hat R$ operators are satisfied, hence
\begin{eqnarray}
(\sum_{i=1}^{N_c} \sum_{z=1,2} \hat A_{i,z} \hat A^{\dagger}_{i,z})|\Psi^V_3\rangle =0,
\quad \hat P|\Psi^V_3\rangle =0, \quad \hat R|\Psi^V_3\rangle =0.
\label{EQ34}
\end{eqnarray}
For other possibilities to find a checkerboard type of phase, see
Ref.(\cite{VVX8}). For the presently used Hamiltonian in (\ref{EQ26}), and
concentration domain, the choice in (\ref{EQ33}) is unique.

Hence the ground state at $V > 0$ and $N=3.5 N_c$ is $|\Psi^V_g(3.5N_c)\rangle =
|\Psi^V_3\rangle$, where
\begin{eqnarray}
  |\Psi^V_g(3.5N_c)\rangle =(\prod_{i \in A} [\hat d^{\dagger}_{i,\uparrow} +
\hat d^{\dagger}_{i,\downarrow}])(\prod_{i=1}^{N_c} [\hat d^{\dagger}_{i,\uparrow} +
\hat d^{\dagger}_{i,\downarrow}]) (\prod_{i=1}^{N_c} \hat A^{\dagger}_{i,z=1}
\hat A^{\dagger}_{i,z=2}) |0\rangle, 
\label{EQ35}
\end{eqnarray}
Comparing the $V > 0$ ground state in (\ref{EQ35}) to the $V=0$ ground state
from (\ref{EQ27}), it is seen that the homogeneity in (\ref{EQ27}) disappears
at $V > 0$, different sites behave differently, and again a CDW type of
behavior appears.

\section{Summary and Discussions}

Since the effect of the nearest neighbor Coulomb repulsion V often seems
antagonistic, and the spectrum of phases in which  the V term
can be present is extremely
broad, the aim of the present paper is to understand those effects of V which
are intimately related to V itself, and not to the phase characteristics on
which it acts. On this line exact ground states are presented, which
demonstrate that $V > 0$ in the large concentration regime destroys the phase
homogeneity, the effect being missing in the low concentration domain. Since
$V > 0$ means usually non-integrability, the presented exact results are
deduced for non-integrable systems (the diamond chain, and the two dimensional
two bands system with many-body spin-orbit interaction) using a technique based
on positive semidefinite operator properties. In the light of the obtained
results, the antagonistic behavior appears because the actions in two different
concentration regions are compared, while the resilience to V appears outside
of the high concentration regime.


ACKNOWLEDGEMENT:

Research supported by Chinese Academy of Sciences President's International
Fellowship Initiative, PIFI Grant No: 2025PVA0087.

E-MAIL ADDRESS OF THE AUTHOR:

gulacsi@phys.unideb.hu

DATA AVAILABILITY STATEMENT:

The data that support the results are available from the author upon
reasonable request.




\end{document}